\newcommand {\arcsec}{$^{\prime\prime}$}

\newcommand {\hMpc}{$h^{-1}$Mpc}

\newcommand {\ds}{\displaystyle}
\newcommand {\gtrsim}{{\;{\ds\lower.7ex\hbox{$>$}\atop\ds\sim}\;}}
\newcommand {\lesssim}{{\;{\ds\lower.7ex\hbox{$<$}\atop\ds\sim}\;}}

\newcommand {\qso}{1335.8+2834}

\documentstyle[12pt,aasms4]{article}

\begin{document}

\lefthead{Yamada et al.}
\righthead{Red galaxies near a radio-loud quasar at $z=1.1$}

\title{CLUSTERING OF RED GALAXIES NEAR THE RADIO-LOUD QUASAR 
1335.8$+$2834 AT z$\,$=$\,$1.1 
\footnote{Partly based on observations made with the University of
Hawaii 2.2$\,$m telescope, and the Isaac Newton Telescope, operated on the
island of La Palma by the Royal Greenwich Observatory in the Spanish
Observatorio del Roque de los Muchachos of the Instituto de Astrof\'\i
sica de Canarias. }}
\author{Toru Yamada, Ichi Tanaka}
\affil{Astronomical Institute, Tohoku University, Aoba-ku, Sendai 980-77, 
Japan}
\author{Alfonso Arag\'on-Salamanca, Tadayuki Kodama}
\affil{Institute of Astronomy, Madingley Road, Cambridge, CB3 0HA, UK}
\author{Kouji Ohta}
\affil{Department of Astronomy, Kyoto University, Kyoto, 606-01, Japan}
\and
\author{Nobuo Arimoto}
\affil{Institute of Astronomy, University of Tokyo, Mitaka, Tokyo 181, Japan}
\begin{abstract}
We have obtained new deep optical and near-infrared images of the field
of the {\it radio-loud\/} quasar 1335.8$+$2834 at $z=1.086$ where an
excess in the surface number density of galaxies was reported by
Hutchings et al.  [AJ, 106, 1324] from optical data. We found a
significant clustering of objects with very red optical-near infrared
colors, $4 \lesssim R-K \lesssim 6$ and $3 \lesssim I-K \lesssim 5$
near the quasar.  The colors and magnitudes of the reddest objects are
consistent with those of old (12 Gyr old at $z=0$) passively-evolving
elliptical galaxies seen at $z=1.1$, clearly defining a `red envelope'
like that found in galaxy clusters at similar or lower redshifts.  This
evidence strongly suggests that the quasar resides in a moderately-rich
cluster of galaxies (richness-class $\geq 0$). There is also a
relatively large fraction of objects with moderately red colors ($3.5 <
R-K < 4.5$) which have a distribution on the sky similar to that of the
reddest objects. They may be interpreted as cluster galaxies with some
recent or on-going star formation.
\end{abstract}


\section{Introduction}

Clusters of galaxies from low to high redshift provide a unique
opportunity to trace back the evolution of galaxies in dense
environments. Detailed studies of elliptical galaxies in nearby and
intermediate redshift clusters showed that their photometric properties
are homogeneous and evolve only mildly with redshift at $z \lesssim
1$.  This is consistent with the so-called `passive evolution', namely
the change of the photometric properties of elliptical galaxies by the
aging of their stars only, with a formation epoch of at least $\sim 10$
Gyr ago (Bower, Lucey \& Ellis 1992; Arag\'on-Salamanca et al. 1993;
Ellis et al. 1997; Dickinson 1997b; Stanford, Eisenhardt, \& Dickinson
1997). On the other hand, some S0 and spiral galaxies seem to
experience faster evolution (e.g., Dressler et al. 1994; Dressler and
Smail 1997).   It is clear that pushing these studies to higher
redshifts should provide stronger constraints on the epoch of cluster
galaxy formation.  Unfortunately, only a very limited number of
clusters and cluster candidates are known beyond $z \simeq 1$
(Dickinson 1995; 1997b), and thus the identification of high-redshift
clusters itself should be the next important step.

Near-infrared (NIR) observations are essential in identifying clusters of
galaxies at z $\gtrsim 1$.  It is known that old passively-evolving
galaxies observed at high redshifts have very red optical-NIR colors
(Arag\'on-Salamanca et al. 1993; Dickinson 1995; Yamada and Arimoto
1995), since their optical flux rapidly decreases as the
4000$\,$\AA\ break shifts toward longer wavelengths.  For example, a
3$\,$Gyr-old passively-evolving model galaxy observed at $z = 1$ has
$I-K\sim 4$ which is redder than most galactic stars and most field
galaxies.  Therefore by selecting objects with red optical-NIR colors,
the contrast between cluster and field galaxies will be very much
enhanced if clusters are dominated by old galaxies.

It has been known for some time that radio-loud AGNs (both
radio-galaxies and quasars) are frequently located in cluster
environments (see, e.g., Yee \& Green 1987; Ellingson, Yee \& Green
1991; Dickinson 1995). Thus, targeted searches in the fields of high
redshift radio-loud AGN provide an efficient approach to the problem of
finding the (probably) rare galaxy clusters at $z>1$ (see Dickinson
1997c for a review).

In this {\it Letter\/} we report the clustering of very red
objects in the vicinity of a quasar at $z = 1.1$ where some excess in
the surface number density of galaxies had already been reported by
Hutchings et al.  (1993) based on their optical data. We obtained 
deeper $R$ and $I$ optical images and NIR $K$-band data of the
region and confirmed the clustering of galaxies near the quasar; their
optical-NIR colors and magnitudes are consistent with bright
cluster galaxies formed at higher redshift and observed at $z = 1.1$.
%Section~2 describes the new data. 
%The sky distribution and color properties of the objects are discussed in
%section 3. In section 4 we discuss the properties of the cluster
%galaxies and the quasar environment. 
 The cosmological parameters
$H_0 = 50\,$km$\,$s$^{-1}\,$Mpc$^{-1}$ and $q_0 = 0.5$ have been
assumed.

\section{Observations and data reduction}

With the aim of identifying high redshift cluster galaxies with red
optical-NIR colors, we have targeted the field around the
radio-loud
\footnote{Hutchings et al. (1993) classified this quasar as
radio quiet.  However, a plausible radio counterpart with $S_{\rm
6cm}=55$ mJy exists at ($\alpha_{1950}, \delta_{1950}) =
(13^h\,35^m\,47^s.8\pm1.2, +28^\circ\,20^\prime\,19^{\prime\prime}\pm
22)$ (Gregory and Condon 1991), while the quasar locates at
$(13^h\,35^m\,48^s.39, +28^\circ\,20^\prime\,24^{\prime\prime}.3)$
(V\'eron-Cetty and V\'eron 1996). We estimate that the logarithmic
radio-optical luminosity ratio of the quasar is
$\log(L_\nu^{6cm}/L_\nu^{V}) = 3.34$. Thus the \qso\ should be
classified as radio-loud. In the recent VLA FIRST survey catalog (White
et al. 1997), where the typical positional error is less than
$1\,$arcsec, there is a radio source at $(13^h\,35^m\,48^s.409,
+28^\circ\,20^\prime\,22^{\prime\prime}.61)$.} 
quasar 1335.8$+$2834 at
$z = 1.086$. Hutchings, Crampton \& Persam (1993) found that the
surface density of galaxies within $\sim 100$\arcsec\ separation ($\sim
1\,$Mpc at $z = 1.1$) from the quasar is three times larger than the
average density of their control fields.  Using narrow-band images
centered on the [OII]$\lambda 3727$\AA\ line redshifted to $z=1.086$,
they also found several emission-line galaxies near the quasar,
suggesting the excess density is associated with the quasar redshift.

Optical images were obtained with the Isaac Newton Telescope at la
Palma on the 28th of February, 1995. A TEK $1024\times1024$ CCD with a
projected pixel size of $0.59\,$arcsec was used. The weather conditions
were mostly photometric. Total exposures of $8100\,$sec in the $R$ band
and $7500\,$sec in the $I$ band were gathered, subdivided in $900$ or
$600\,$sec sub-exposures. The FWHM sizes of the stars were typically
1.5 ($R$) and 1.8 ($I$) arcsec.  The frames were  bias-subtracted,
flat-fielded and median-stacked in the usual manner.

The $K$-band images were obtained with the University of Hawaii 2.2m
telescope equipped with the QUIRC camera on the 29th of February,
1996.  The detector was the  $1024\times1024$ HgCdTe array with a
projected pixel size of $0.18\,$arcsec. Since the field of view of this
camera ($3 \times 3\,$arcmin$^2$) is smaller than that of the optical
CCD ($7\times7\,$arcmin$^2$), we covered only the field which contains
the region where the surface density had been evaluated to be high from
the optical data.  In this letter, we concentrate mainly on the
properties of the objects contained within the $K$-band field.  The net
exposure time of the $K$-band image is $6480\,$sec, subdivided in many
$180\,$sec dis-registered exposures.  The FWHM size of the stellar
images was $0.9\,$arcsec. Flat-field frames were built using a running
median of $\simeq10$ images in a time sequence.  The flat-fielded
images were then median-stacked and a constant sky value was
subtracted.  The data reduction was done with IRAF.

Object detection was carried out using FOCAS (Jarvis and Tyson 1981)
with a detection threshold of $2.5\sigma$ of the sky fluctuations per
pixel. A minimum of $\sim (FWHM)^2$ connecting pixels above the
threshold was required for detection.  Landolt (1992) standards in the
$R$ and $I$ bands, and UKIRT bright standards in $K$ were used for 
photometric calibration, yielding  zero-point errors
smaller than $0.1\,$mag.  We measured the colors of the objects within
a fixed aperture with a diameter of $3.6\,$arcsec, twice the FWHM
seeing in the $I$ band image (the worse case), after correcting the
seeing differences among the images. The APPHOT task in the IRAF was
used for the photometry.  Total magnitudes were also estimated using
template growth curves determined from 10 isolated galaxies in the
field.

The source detection is complete for objects brighter than 24.5, 23.5,
and 20.0 magnitudes in $R$, $I$, and $K$ respectively.  These limits
were estimated from the turn-off of the number counts in the `field
region' (western half of the $K$-band image), which avoids the region
of the cluster candidate (see ahead).  The counts are consistent,
within  $\simeq1\sigma$, with published field number counts (e.g.,
Smail et al. 1995; Djorgovski et al. 1995).  The internal photometric
errors in the aperture magnitudes were evaluated using the photon count
statistics and also by changing the sky regions in the local sky
subtraction procedure. Typical errors are $\sim 0.1$ mag in $R$ and
$\sim 0.15$ mag in $I$ and $K$ at the completeness limits.  The error
in the $K$-band total magnitude (used in the color-magnitude diagram
below) is estimated to be $\sim 0.35$ mag at $K \sim 20\,$mag, where
the uncertainties in our growth-curve-fitting procedure have been taken
into account.

\section{The spatial distribution and colors of the galaxies}

Figure 1 shows the three-color image of the $K$-band field, and
figure~2a shows the distribution of the objects detected in the
$K$-band image. Objects with different colors are shown as different
symbols. An excess in the number density of faint objects near the
quasar (denoted by `QSO') is seen.  Hutchings et al.  (1993) divided
the image into cells and found a $3.5\sigma$ surface density excess in
the number of objects in a $100\times100\,$arcsec$^2$ region centered
on the quasar.  We made a similar analysis on our $R$ and $I$ optical
images and found the excess surface density in a similar region to be
$3 \sigma$ above the counts of the surrounding field.  Moreover,
red objects are seen much more frequently in the eastern (left) half of
the image near the quasar.  The color of most objects in the western
side of the field is $R-K \lesssim 3.5$ (see below).  It is clear that
objects with red optical-NIR colors are responsible for most of the
excess in the number density.  Since stars earlier than M3V have colors
$R-K \lesssim 3.5$ (Johnson 1966; Bessell 1990), most of these red
objects are probably galaxies (see figure~4). An un-evolved elliptical
galaxy (the reddest case for a galaxy without extinction) has $R-K \sim
3.5$ at $z\sim 0.4$, thus these red objects are probably galaxies at
higher redshifts.

A fairly bright galaxy near the quasar (G1 in Figures~1 and~2) has $K =
17.3$, $R-K = 5.7$, and $I-K = 4.3$, which agree very well with the
expectations for a brightest cluster galaxy at $z \sim 1.1$
(Arag\'on-Salamanca et al 1993).  A closer look at the spatial
distribution of the overdensity indicates that the contrast is
maximised when G1 is taken as the center, instead of the quasar.
Figure~2b shows the surface-density-contrast profile, centered on G1,
of all the objects detected in the $K$ band as well as those with $R-K
> 3.5$.  The excess surface density of $K$-band detected objects in a
100 arcsec box centered on G1 is  $6\sigma$ above the counts of the
surrounding field, which is significantly above that found in the
optical images.  Moreover, if we consider only the red objects, the
contrast of the putative cluster is significantly enhanced to
$\simeq10\sigma$ above the field counts, which clearly indicates that
the use of optical-NIR colors is a powerful technique to identify
high-redshift clusters.

\vskip -0.02cm

The size of the region with the excess density is $\sim 1$ arcmin in
diameter, corresponding to $\sim 0.5\,$Mpc at $z = 1.1$, which is
consistent with that expected for a rich galaxy cluster. If we count
all the galaxies with $R-K > 3.5$ as
cluster members, nearly 30 galaxies remain after correcting for field
contamination in this color range.  This roughly
corresponds to the Abell richness class~0.  This could be a
lower-limit for the cluster richness since we have only counted the red
objects as cluster members. A more accurate estimate of the richness is
not possible without galaxy redshifts.

Figure 3 shows the $R-K$ vs $K$ color-magnitude diagram for the objects
in the $K$-band field. We tentatively define the `cluster region' as
the eastern half of the field, and show the objects in this area with
filled circles.  The color distributions in the two regions are very
different. There are 37 galaxies with $3.5 < R-K
< 6$ and $17 < K < 19$ in the eastern `cluster
region' but only 6 such objects in the western `field' region. In the
cluster region, there is a group of objects with similar colors and
magnitudes, $4.5 < R-K < 6$ and $K = 17$--$19\,$mag.  They are consistent
with a homogeneous population of galaxies seen at a similar redshift,
which suggests that the clustering is probably real.  These objects
have $I-K\gtrsim 3$. The $I-K$ color is very redshift-sensitive at $z \sim 1$
due to the fact that the $4000\,$\AA\ break is located near or just
inside the $I$ band at this redshift. Thus, the red $I-K$ colors strongly
suggest that these objects are old galaxies at $z \gtrsim 1$.  For
comparison, on Figure~3 we also plot the model colors and magnitudes of
passively-evolving elliptical galaxies calculated with the models
introduced by Kodama and Arimoto (1997).  The model parameters are
chosen so that the color-magnitude relation at $z=0$ (interpreted as a
mass-metallicity sequence) agrees with that of the Coma cluster
(Bower et al. 1992).  The color and luminosity evolution
predicted by such models agrees very well with the observed one for
cluster ellipticals (see Kodama and Arimoto 1997). The solid lines
indicate the expected colors of passively-evolving model galaxies with
$M_V = -22.0$ and $-18.5$ mag at $z=0$, which are formed at $z_{f}=4.5$
($t_{age} = 12\,$Gyr at $z=0$).  Tilted straight lines indicate the
expected color-magnitude relation at different redshifts (see caption). 
The colors of the reddest galaxies in the complete
sample are consistent with passively-evolving old galaxies observed at
$z=1.1$. They define a `red envelope' as expected if they are the
oldest cluster galaxies at this redshift, although a clear
color-magnitude sequence is not seen.

\vskip -0.02cm

The color-magnitude diagram also reveals a significant fraction of
objects with $3.5 < R-K < 4.5$ and $K \sim 17-19$.  Since their color
distribution seems somewhat separated from the redder galaxies, it is
possible that these objects belong to a different system at lower
redshift ($z\sim 0.6$). However, the sky
distribution of these moderately red objects is similar to that of the
reddest objects (cf. Figure~2a), so it is possible that they are in the
same system. Without redshifts, these to possibilities cannot be discriminated.

Figure 4 shows the two-color diagram for the `cluster' (west half) and
`field' (east half) galaxies. The solid line shows the colors of
elliptical galaxies observed at various redshifts derived from the same
models as in Figure~3, representing the expected locus of the reddest
galaxies observed at any redshift in the absence of reddening.  It is
clear that the reddest galaxies in both $R-K$ and $R-I$ are compatible
with being passively-evolving cluster ellipticals at $z=1.1$.  On the
other hand, there is a number of galaxies with moderately red $R-K$
color ($3.5 \lesssim R-K \lesssim 5$) and bluer $R-I$ colors.  Hutchings et
al. (1993) had also noticed the bimodal distribution of $R-I$ colors in
the excess-density region, which is not seen in their control field.
If they are cluster members, their bluer optical colors would suggest a
certain amount of current or recent star forming activity.

\section{Discussion and Conclusions}

We have found significant clustering of galaxies with very red
optical-NIR colors near the radio-loud quasar 1335.8$+$2834 at
$z=1.1$.  The reddest objects ($4.5 \lesssim R-K  \lesssim 6$) are
responsible for nearly half of the surface density excess, and their
colors and magnitudes are consistent with those of passively-evolving
old galaxies seen at $z\sim 1.1$.  If they are at the quasar redshift,
galaxy evolution models suggest that in the absence of reddening the
reddest objects are already $\sim 2$--$4\,$Gyr old.  The range of
colors displayed by the galaxies can be readily explained by the
presence of some field galaxies at a variety of redshifts and by
cluster galaxies with a variety of star formation histories such as
disk galaxies formed a few Gyr before $z=1.1$, or old galaxies with
young bursts of star formation.  Redshifts and morphologies are needed
to study the evolutionary properties of the galaxies in detail.

Other examples of clustering of red galaxies near a radio-loud object
at comparable redshifts have been found by Dickinson and his
collaborators (Dickinson 1995; 1997a). They observed the fields of
powerful radio galaxies at intermediate and high redshifts with a
similar optical and NIR imaging strategy.  The cluster around 3C~324 is
their most outstanding case. Redshifts for a few hundred objects
confirmed a cluster at $z=1.21$ associated with 3C~324, although many
interlopers in another group or cluster at $z=1.15$ were also found. A
similar contamination by a group or cluster at a slightly
different redshift cannot be excluded in the case of 1335.8$+$2834
without extensive redshift data.

\vskip -0.095cm

Hintzen, Romanishin \& Valdes (1991) have also found significant
clustering of galaxies around  $z = 0.9$--$1.5$  radio-loud quasars.
The galaxies responsible for the excess they detected seemed to be more
luminous in the observed $R$-band than present-day brightest cluster
galaxies.  However, the galaxies responsible for the excess we have
found in the field of 1335.8$+$2834 do not seem to be over-luminous in
the $K$-band.  Since the observed $R$-band samples ultraviolet light at
these redshifts, evolutionary effects such as the presence of moderate
amounts of young stars can boost up the observed flux considerably,
while the NIR should show significantly milder evolutionary effects.

%It
%would be interesting to see whether the galaxies responsible for the
%excess found by Hintzen et al.  are also over-luminous in the NIR,
%where evolutionary effects are expected to be significantly milder.

\vskip -0.095cm

Excess in the galaxy number density has also been found for
higher-redshift quasars or radio galaxies (Dressler et al. 1993, 1994;
Hutchings 1995; Pascarelle et al. 1996). These authors claim that the
excess galaxies are small compact objects which may be the `building
blocks' of today's luminous galaxies.  In contrast, many of the
galaxies found near 1335.8$+$2834 are fairly luminous ($\gtrsim
0.5L_*$).  Similarly, an excess of $\sim L_*$ galaxies was found near
radio-loud quasars at $z\sim2$ by Arag\'on-Salamanca, Ellis, and O'Brien
(1996). The situation is therefore not clear, and a systematic study of
the fields of a large well-defined sample of known high redshift
objects is clearly needed.

\vskip -0.1cm

In summary, we have shown that the combination of deep optical and NIR
imaging is a very powerful tool in revealing high redshift clusters of
galaxies and/or their progenitors. Such clusters can provide samples of
high redshift galaxies for evolutionary studies at large look-back
times, which can greatly contribute to our knowledge of galaxy
formation and evolution at very early epochs. While this kind of
study is very successful at finding candidate clusters, redshifts
and morphologies will be ultimately needed.

\acknowledgements

\noindent This work was partially supported by a grant-in-aid for
Scientific Research of the Japanese Ministry of Education, Science,
Sports and Culture (No.07041104  and No.08740181). Part of this work
was also supported by the Foundation for the Promotion of Astronomy of
Japan. TY was a Special Post-Doctoral Researcher of the Institute of
Physical and Chemical Research (RIKEN) and a part of this work was
supported by this institute.  AAS acknowledges generous financial
support from the Royal Society and the Particle Physics and Astronomy
Research Council.  TK thanks to the Japan Society for the Promotion of
Science Postdoctoral Fellowships for Research Abroad. KO was a visiting
astronomer of the Institute for Astronomy, University of Hawaii and
thanks their hospitality during his stay.

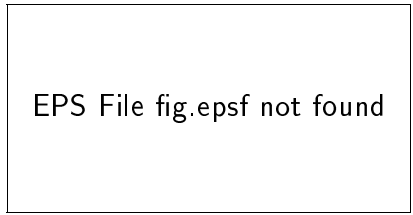
\figcaption{$RIK$ three color image of the field near the radio-loud
quasar 1335.8$+$2834 at $z=1.086$ (QSO). North is up, East is left. 
The field is 160 arcsec on each side, which
corresponds to $0.62\,$\hMpc\ for $q_0$=0.5.\label{fig-1}}

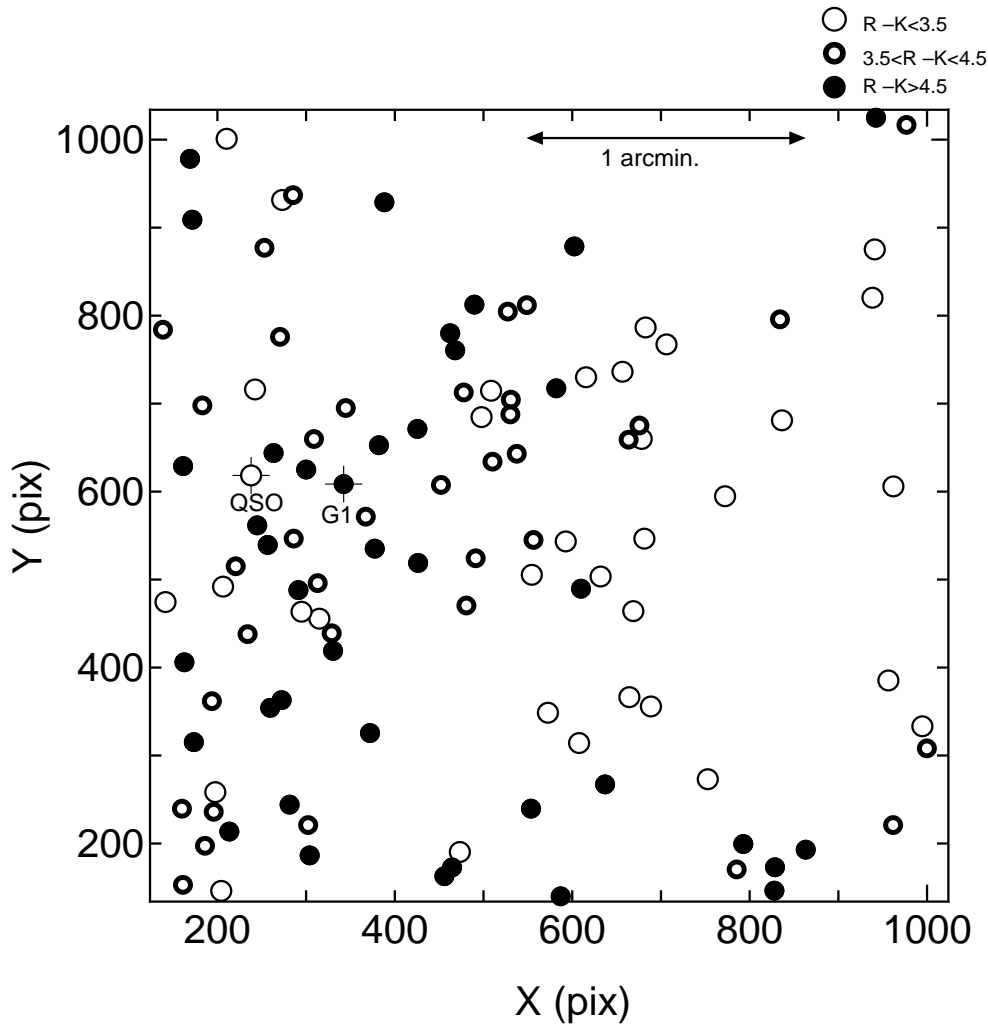
\figcaption{ {\bf (a)} Distribution of the $K$-selected objects on the
sky.  Scale and orientation as in Figure 1. Objects in different color
ranges are shown as different symbols.  {\bf (b)}
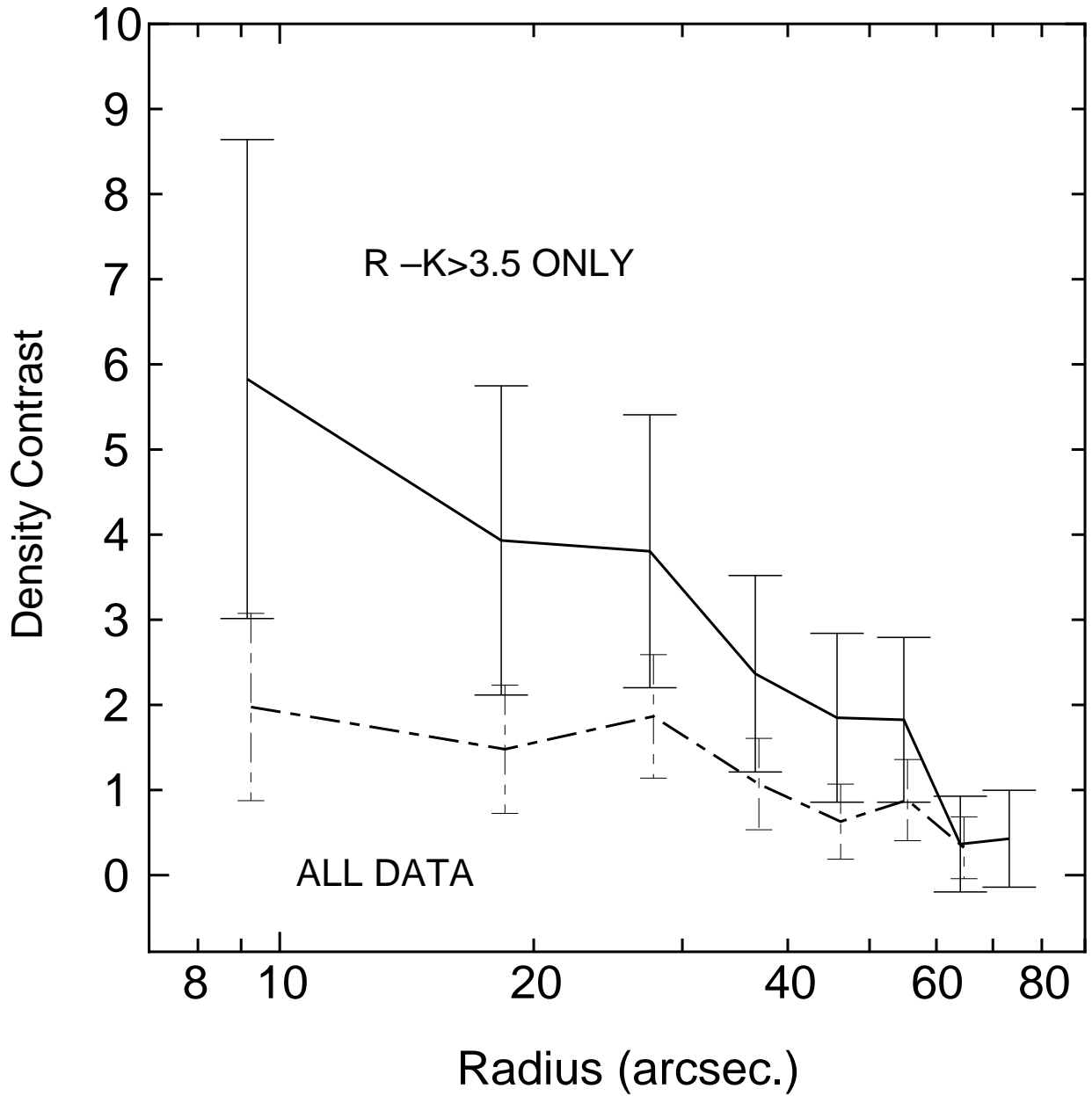
Surface-density-contrast profile, ($\Sigma(\theta)-\langle\Sigma
\rangle)/\langle\Sigma \rangle$, for all the $K$-selected objects
centered on galaxy G1 (dash-dotted line) and for the objects with
$R-K\geq 3.5$ (solid line).  The average field number densities are the
values obtained at the western half of the image avoiding the cluster
region.  \label{fig-2}}

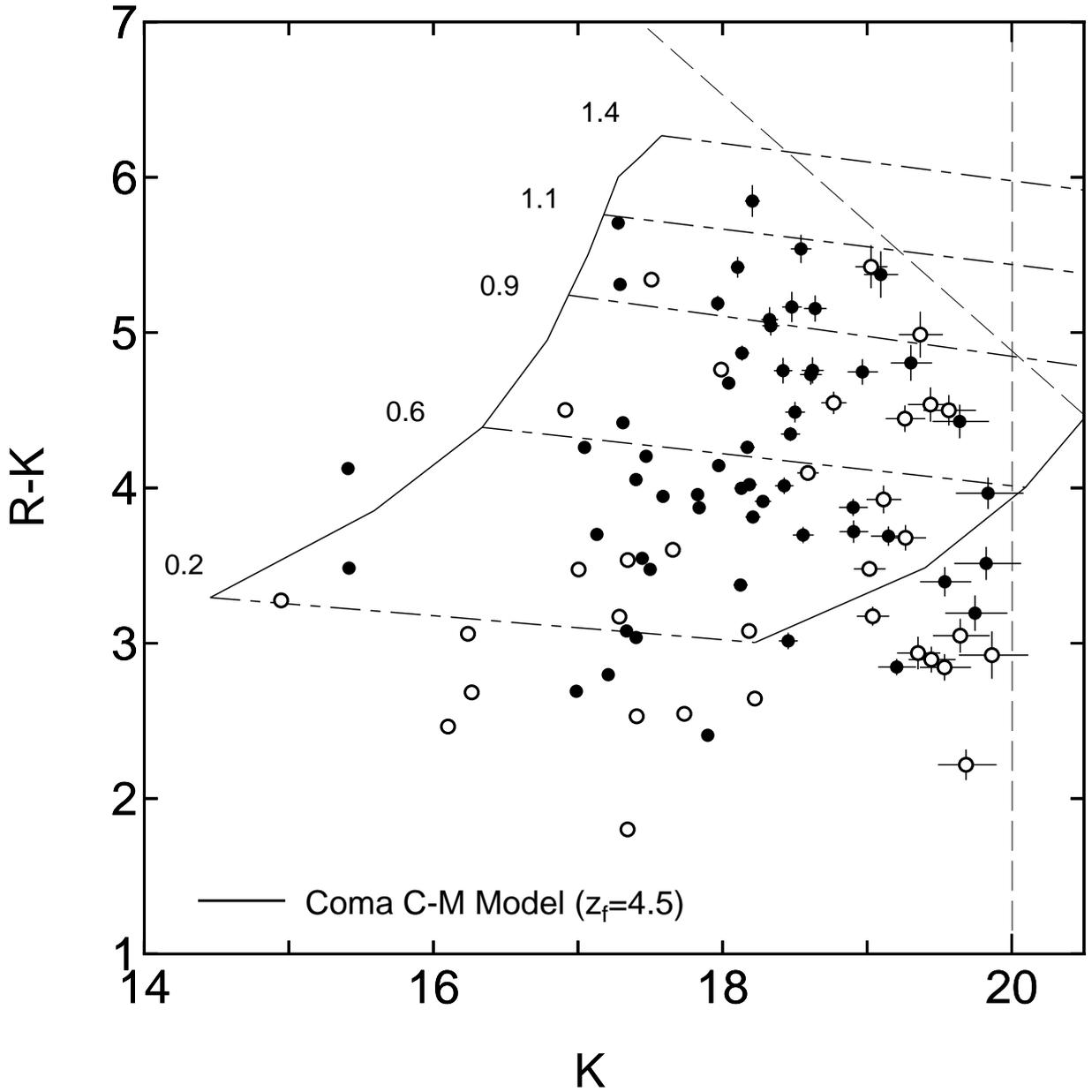
\figcaption{$R-K$ vs. $K$ color-magnitude diagram of the complete
photometric sample. The completeness limits are shown by the dashed
lines.  The objects in the putative cluster region, namely the eastern
half of the frame, are shown as filled circles. Those in the western
half are shown as open circles. The predicted colors and magnitudes of
model galaxies with $M_{V} = -22.0$ and $-18.5$ mag at $z=0$, formed at
$z_f = 4.5$ are shown as solid lines.  Tilted dash-dotted lines indicate the
expected color-magnitude relation observed at $z = 0.2$, 0.6, 0.9, 1.1,
and 1.4, which correspond to galaxy ages of 8.9, 5.4, 3.9, 3.2, and 2.5
Gyr, respectively  (computed from the models of Kodama and Arimoto
1997).  \label{fig-3}}

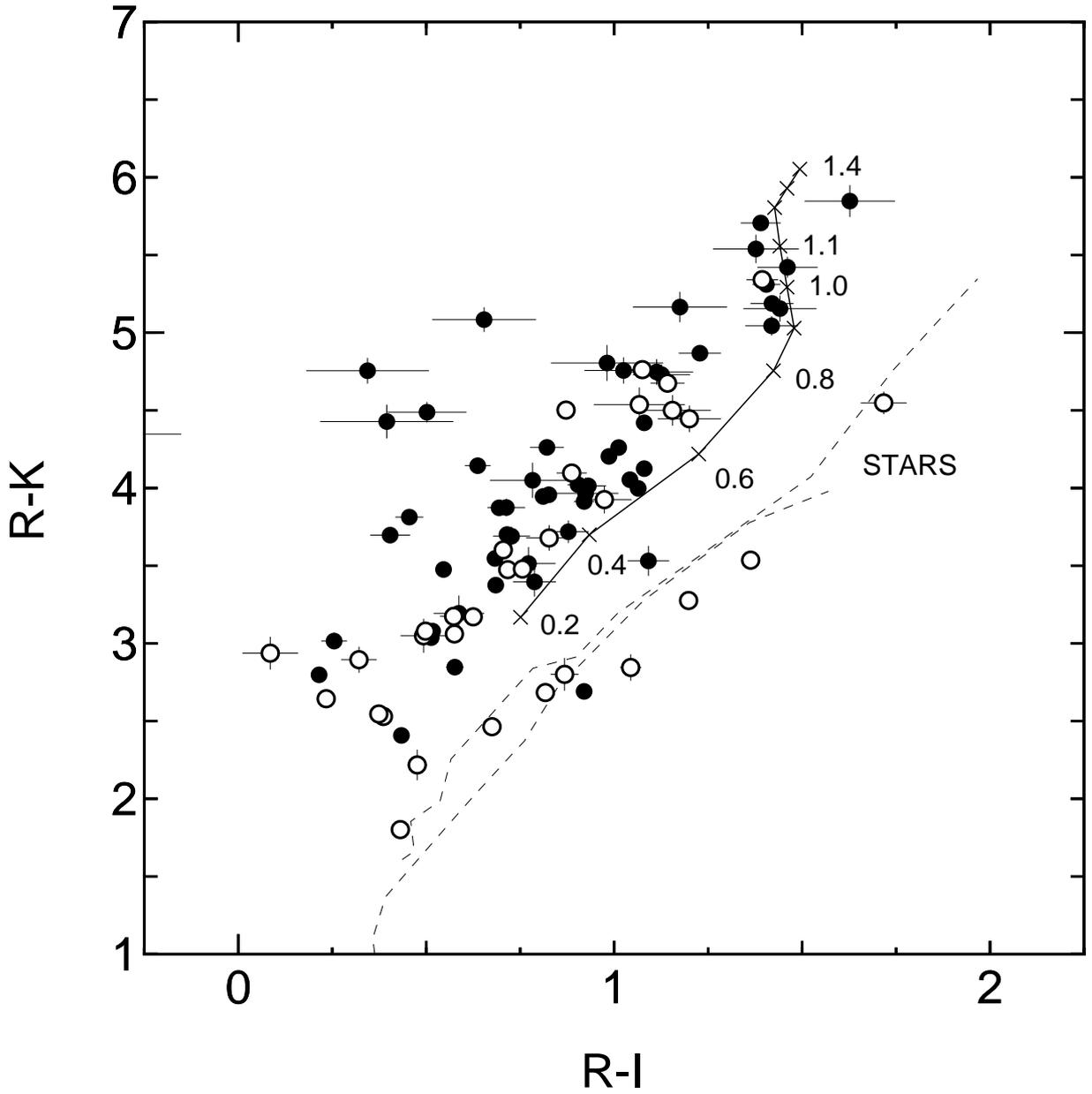
\figcaption{$R-K$ vs. $R-I$ color-color diagram for the complete
sample. Symbols are as in Figure~3. The solid line indicates the
predicted colors of model galaxies with $M_V = -20.5$ mag at $z=0$ observed
at various redshifts derived from the same models as in Figure~3.  The
labeled crosses indicate the observed redshifts.  Colors of dwarf
(G0V--M5V) and giant (G5III--M6III) stars (Johnson 1966; Bessell 1990) are
also plotted for reference (dashed lines). \label{fig-4}} 

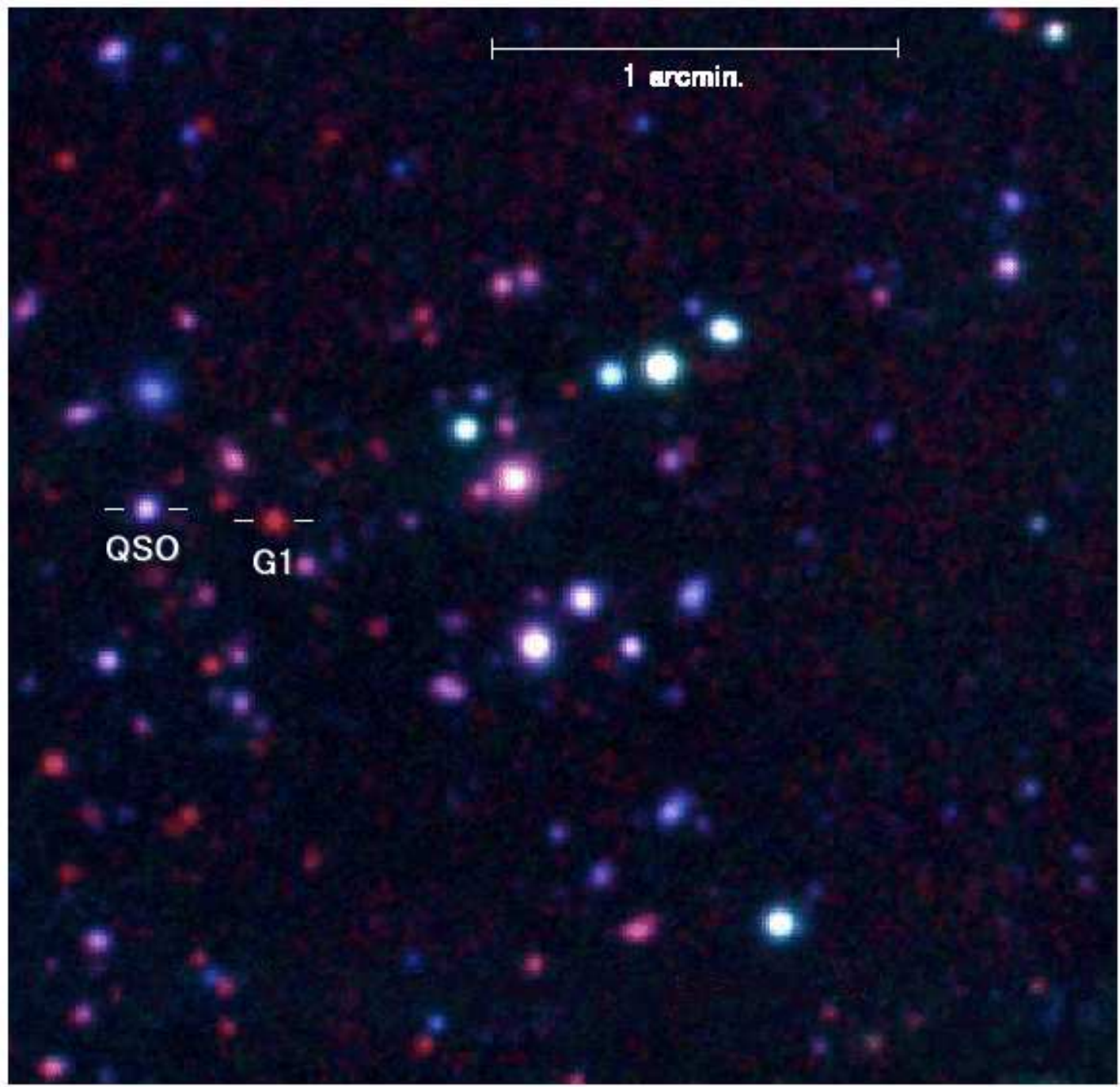
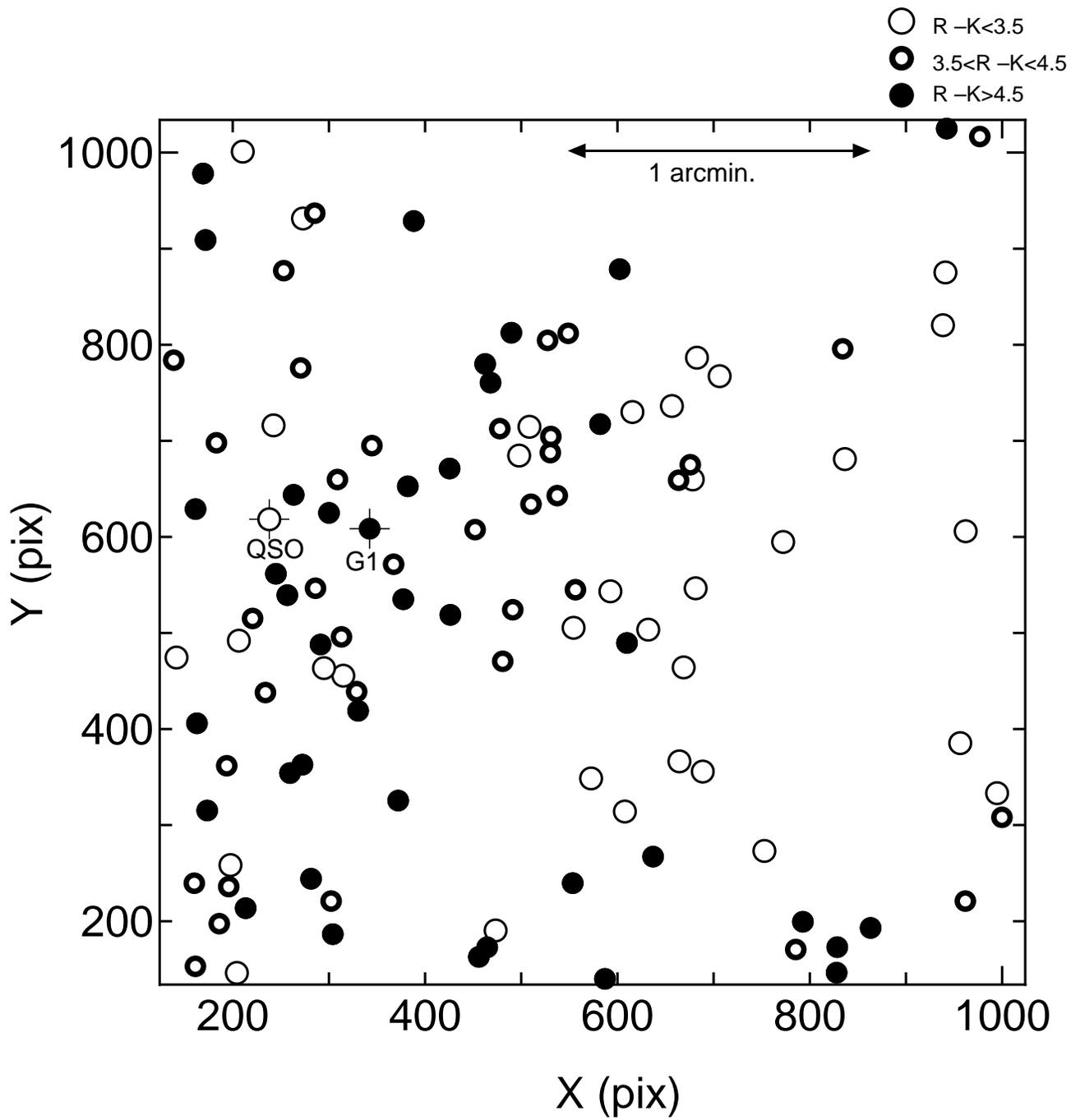
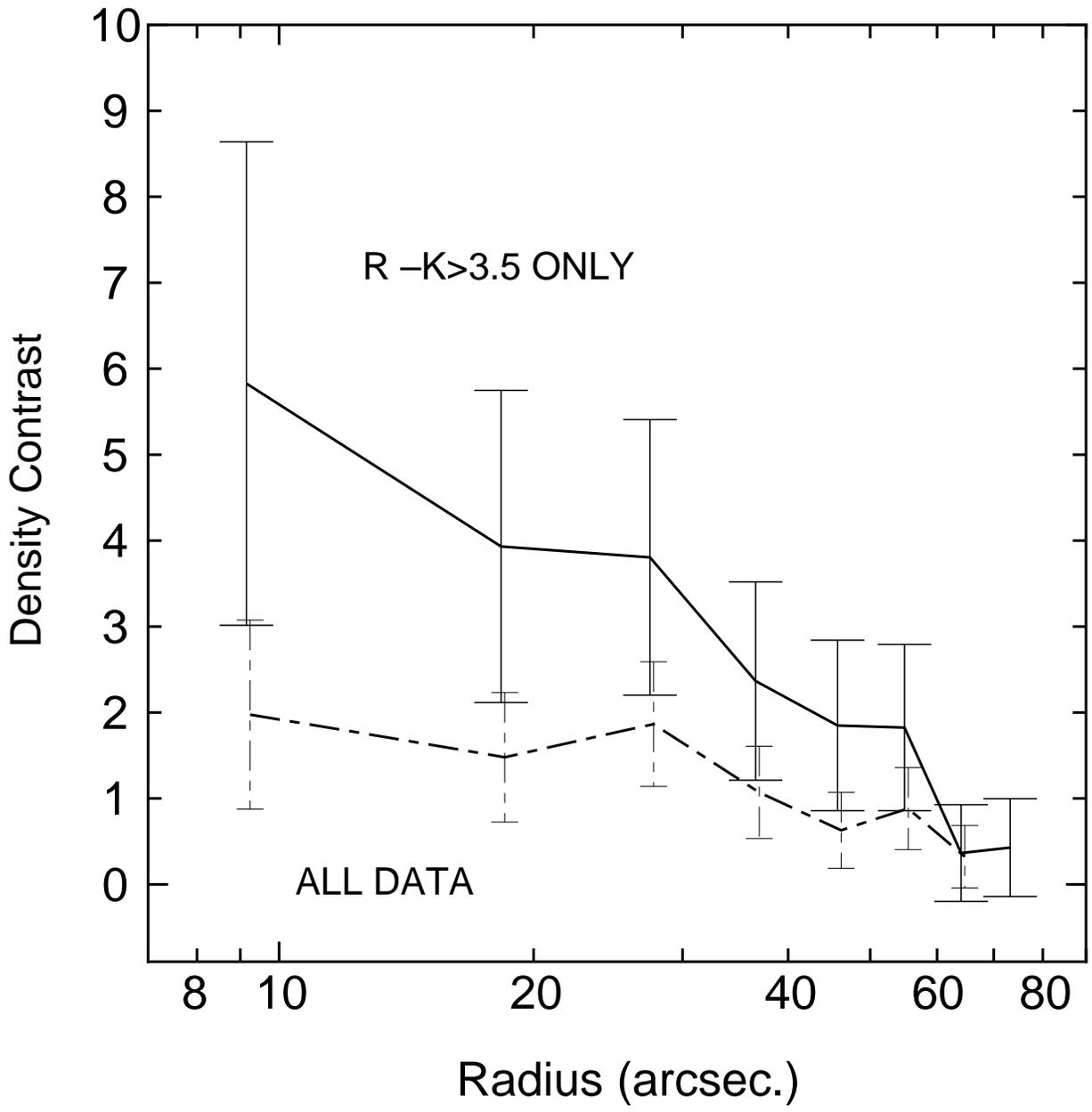
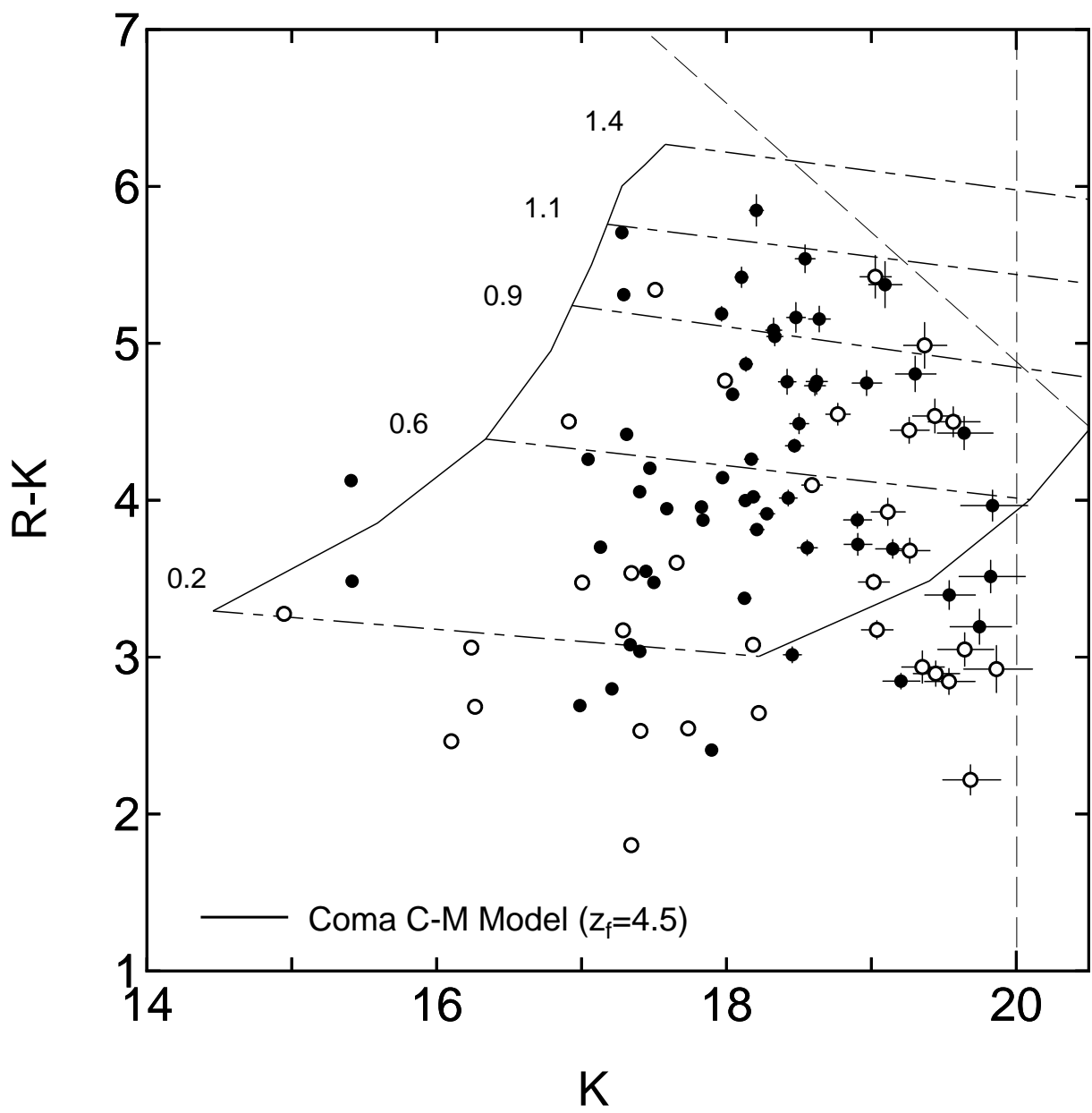
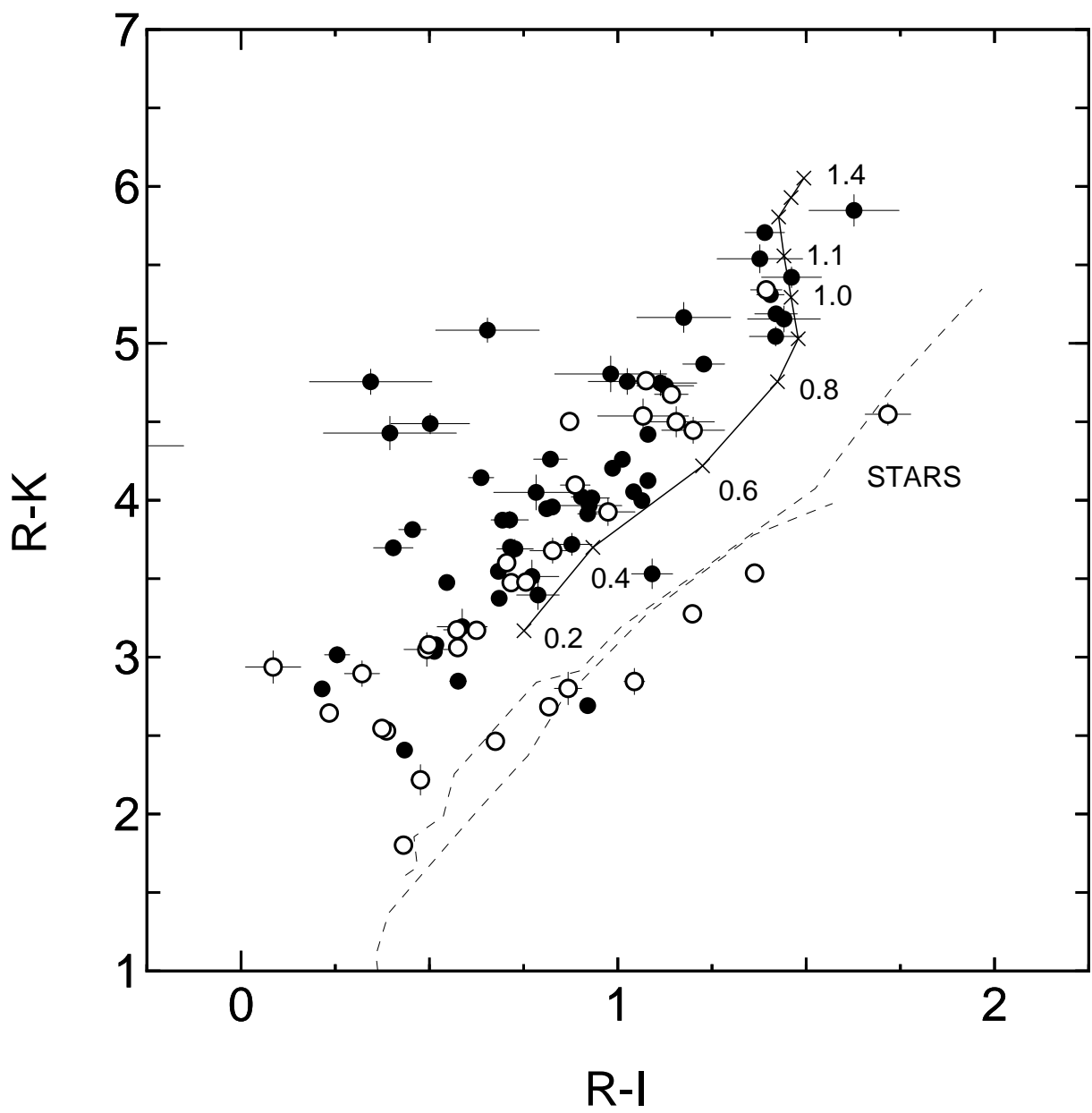
\end{document}